\documentclass[10pt, conference, compsocconf]{IEEEtran}
\usepackage[pdftex]{graphicx}

\usepackage[cmex10]{amsmath}

\usepackage[caption=false,font=footnotesize]{subfig}

\usepackage{fixltx2e}
\usepackage{url}
\begin{document}
\title{Agent-based Versus Macroscopic Modeling of Competition and Business Processes in Economics}

\author{\IEEEauthorblockN{Valentas Daniunas}
\IEEEauthorblockA{\\
Institute of Lithuanian Scientific Society\\
Vilnius, Lithuania\\
mokslasplius@itpa.lt}
\and
\IEEEauthorblockN{Vygintas Gontis, Aleksejus Kononovicius}
\IEEEauthorblockA{Institute of Theoretical Physics and Astronomy\\
Vilnius University\\
Vilnius, Lithuania\\
vygintas@gontis.eu, aleksejus.kononovicius@gmail.com}
}

\maketitle

\begin{abstract}
Simulation serves as a third way of doing science, in contrast to
both induction and deduction. The web based modeling may
considerably facilitate the execution of simulations by other
people. We present examples of agent-based and stochastic models
of competition and business processes in economics. We start from
as simple as possible models, which have microscopic, agent-based,
versions and macroscopic treatment in behavior. Microscopic and
macroscopic versions of herding model proposed by Kirman and Bass
diffusion of new products are considered in this contribution as
two basic ideas.
\end{abstract}


\section{Introduction}
Statistically reasonable models of social systems, first of all
stochastic and agent based, are of great interest for wide
community of interdisciplinary researchers dealing with diversity
of complex systems \cite{Waldrop1992}. Computer modeling
serves as a technique in the for finding relation between micro
level interactions of agents and macro dynamics of the whole
system. Nevertheless, some general theories or methods that are
well developed in the natural and physical sciences can be helpful
in the development of consistent micro and macro modeling of
complex systems \cite{Waldrop1992}. Our own modeling of financial
markets by the nonlinear stochastic differential equations is
based on the empirical analysis of financial data and power law
statistics of proposed equations \cite{GontisScyio2010}. Reasoning
of proposed equations by the microscopic interactions of traders
(agents) looks as a tough task for such complex system. Apparently
the development of macroscopic descriptions for the well
established agent based models would be more consistent approach
in the analysis of micro and macro correspondence. For such
analysis one should select simple enough agent based models with
established or expected corresponding macroscopic description. In
this contribution we discuss few examples of agent based modeling
in business and finance with corresponding macroscopic description
of selected systems.

Kirman's ant colony model \cite{Kirman1993} is agent-based model,
which explains the importance of herding and individuality inside
the ant colonies. As human crowd behavior is ideologically very
similar, this model can be applied to and actually was built as
framework for financial market modeling \cite{Kirman1993,
Alfarano2005, Alfarano2008}. On our website, \cite{RizikosFizika},
we have presented interactive realizations of the original
Kirman's agent-based model (see \cite{RizikosFizikaKirmanAnts})
and of it's stochastic treatment by Alfarano et al.
\cite{Alfarano2005} (see
\cite{RizikosFizikaKirmanAntsStochastic}). Further we follow the
works by Alfarano et al. \cite{Alfarano2005, Alfarano2008} and
introduce our own model modifications in order to obtain more
sufficient agent-based models of financial markets, which would
have an alternative macroscopic description in the terms of
Stochastic Calculus.

Diffusion of new products is a key problem in marketing research.
Bass Diffusion model is a prominent model in diffusion theory
introducing a differential equation for the number of adopters of
the new products \cite{Bass1969}. Such basic macroscopic
description in marketing research can be studied using microscopic
agent-based modeling as well \cite{Mahajan1993}. It is a great
opportunity to explore the correspondence between the two micro
and macro descriptions looking for the conditions under which both
approaches converge. Bass Diffusion model is of great interest for
us as representing very practical and widely accepted area of
business modeling. Web based interactive models, presented on the
site \cite{RizikosFizikaBusiness} serve as an additional research
instrument available for very wide community.

Our web site \cite{RizikosFizika} was setup using WordPress
webloging software. WordPress is user-friendly, powerful and
extensible web publishing platform, which can be adapted to
scientist's needs. There is a wide choice of plugins, which enable
writing of equations (mostly using LaTeX). Though bibliography
management is not as well covered.

Interactive models themselves are independent from WordPress
framework. They were implemented using Java programing language
\cite{Java}, which is better suited for stochastic modeling, and
AnyLogic multi-paradigm simulation software \cite{Anylogic}, which
provides convenient tools to implement agent based models. Either
way by compiling appropriate files one obtains Java applets, which
can be included in to the articles written using WordPress. This
way articles become interactive - user can both theoretically
familiarize himself with the model and test if the claims made in
the article describing model were true. This happens in the same
browser window, thus, transition between theory and modeling
appears to be seemless. As models are implemented as Java applets
all computation occurs on client machine, user must have Java
Runtime Environment installed (it is available free of charge from
Oracle Corp.), and server load stays minimal.

In Sections II and III, we present web-based micro and macro
modeling of selected social systems in more details. Conclusions
and future work are given in the Section IV.

\section{Kirman's model for financial markets and it's stochastic treatment}

There is an interesting phenomenon concerning behavior of ant
colony. It appears that if there are two identical food sources
nearby, ants exploit only one of them at a given time. The
interesting thing is that food source which is in use is not
certain at any point of time. As at some times switches between
food sources occur, though the quality of food sources remains the
same. One could imagine that those different food sources are
different trading strategies or simply actions available to
traders (i.e., buy and sell). Thus, one could argue that
speculation bubbles and crashes in the financial markets are of
similar nature as explotation of food in ant colonies - as quality
of stock and quality of food in the ideal case can be assumed to
be constant. Thus, model \cite{Kirman1993} was created using ideas
obtained from the animal world in order to mimic traders' behavior
in the financial markets.

And actually Kirman, as an economist, developed this model as
rather general framework in context of economic modeling (see
\cite{Kirman1993} and his later bibliography). Though recently his
framework was also used by other authors who are concerned with
the financial market modeling (see \cite{Alfarano2005,
Alfarano2008}). Basing ourselves on the main ideas of these
authors and our previous results in stochastic modeling (see
\cite{GontisScyio2010}) we introduce specific modifications of
Kirman's model providing a class of nonlinear stochastic
differential equations \cite{Ruseckas2010} applicable for the
financial variables.

Original Kirman's one step transition probabilities
\cite{Kirman1993},
\begin{IEEEeqnarray}{l l l}
p( X \rightarrow X+1) &=& (N-X) \left(\sigma_1 + h X \right) , \\
p( X \rightarrow X-1) &=& X \left(\sigma_2 + h [N-X] \right) ,
\end{IEEEeqnarray}
can be rewritten for continuous $x=X/N$ as
\begin{IEEEeqnarray}{l l l}
\pi^{+}(x) &=& (1-x) \left(\frac{\sigma_1}{N} + h x \right) , \\
\pi^{-}(x) &=& x \left(\frac{\sigma_2}{N} + h [1-x] \right) ,
\end{IEEEeqnarray}
where $X$ is a number of agents exploiting chosen trading
strategy, $N$ is a total number of agents in the system. Here the
large number of agents $N$ is assumed to ensure the continuity of
variable $x$, expressing fraction of selected agents, $X$, from
whole population. Note that the transition probabilities depend on
$\sigma_1$, $\sigma_2$ parameters, which govern individual
switches between trading strategies (thus, appropriate terms
depend only on the size of the opposing group), and $h$ parameter,
which governs recruitment (thus, appropriate terms depend on both
sizes - size of the current and opposing groups). Evidently these
probabilities are interrelated
\begin{equation}
p ( X \rightarrow X \pm 1) = N^2 \pi^{\pm}(x) .
\end{equation}
One can write Master equation for the probability density function
of continuous variable $x$ by using one step operators ${\bf E}$
and ${\bf E^{-1}}$ introduced in \cite{vanKampen1992}. Thus,
Master equation can be compactly expressed as
\begin{IEEEeqnarray}{l l}
\partial_t \omega(x, t) =& N^2 \left\{ ({\bf E}-1)[\pi^{-}(x) \omega(x,t)] + \right.\nonumber \\
&\left. +({\bf E^{-1}}-1)[\pi^{+}(x) \omega(x,t)] \right\} .
\end{IEEEeqnarray}
With the Taylor expansion of operators ${\bf E}$ and ${\bf E^{-1}}$ (up to the second term) we arrive at the approximation of the Master equation
\begin{IEEEeqnarray}{l l}
\partial_t \omega(x, t) =& -N \partial_x [\{\pi^{+}(x)-\pi^{-}(x)\} \omega(x,t)] +\nonumber \\
& + \frac{1}{2} \partial^2_x [\{\pi^{+}(x) + \pi^{-}(x)\} \omega(x,t)]  .
\end{IEEEeqnarray}
By introducing custom functions
\begin{IEEEeqnarray}{l l l}
A(x) &=& N \{\pi^{+}(x)-\pi^{-}(x)\} = \sigma_1 (1-x) - \sigma_2 x ,\\
D(x) &=& \pi^{+}(x)+\pi^{-}(x) = 2 h x (1-x) + \nonumber \\
& & +\frac{\sigma_1}{N} (1-x) + \frac{\sigma_2}{N} x ,
\end{IEEEeqnarray}
one can make sure that the above approximation of the Master
equation is actually Fokker-Planck equation (first derived in a
different way in \cite{Alfarano2005})
\begin{equation}
\partial_t \omega(x, t) = - \partial_x [A(x) \omega(x,t)] + \frac{1}{2} \partial^2_x [D(x) \omega(x,t)]  .
\end{equation}
It is known, \cite{Gardiner1997}, that the above Fokker-Planck
equation can be rewritten as Langevin equation (this equation was
also presented in \cite{Alfarano2005})
\begin{IEEEeqnarray}{l l l}
\mathrm{d} x &=& A(x) \mathrm{d} t + \sqrt{D(x)} \mathrm{d} W = \nonumber\\
&=& [ \sigma_1 (1-x) - \sigma_2 x ] \mathrm{d} t + \sqrt{2 h x (1-x)} \mathrm{d} W ,
\end{IEEEeqnarray}
here $W$ stands for Wiener process.

\begin{figure*}[!t]
\centerline{\subfloat{\includegraphics[width=2.5in]{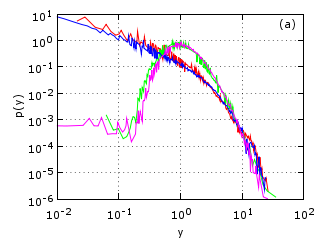}}
\hfil
\subfloat{\includegraphics[width=2.5in]{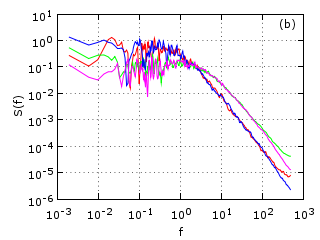}}}
\caption{Agreement between statistical properties of $y$, (a)
probability density function and (b) power spectral density,
obtained from stochastic (blue and magenta curves) and agent-based
(red and green curves) models. Two qualitatively different model
phases are shown: red and blue curves correspond to herding
dominant model phase ($\sigma_1=\sigma_2=0.2$, $h=5$), while green
and magenta curves correspond to individual behavior dominant
model phase ($\sigma_1=\sigma_2=16$, $h=5$).}
\label{fig:sdeAbmKirman}
\end{figure*}

By assuming that market is instantaneously cleared Alfarano et al. \cite{Alfarano2005} have defined return as
\begin{equation}
r = r_0 \frac{x(t)}{1-x(t)} \eta(t) ,
\end{equation}
where $x(t)$ is assumed to be fraction of chartist traders in the
market, while other traders in the market, $1-x(t)$, are assumed
to follow fundamentalist trading strategy, $\eta(t)$ is the change
of chartist mood defined in the same time window as return, in the
most simple case it could be assumed to be a random variable
\cite{Alfarano2005}, and $r_0$ scaling term. Using Ito formula for
variable substitution \cite{Gardiner1997} we obtain nonlinear SDE
for the middle term, $y(t)=\frac{x(t)}{1-x(t)}$,
\begin{equation}
\mathrm{d} y = ( \sigma_1 - y [ \sigma_2 - 2 h] ) (1+y) \mathrm{d} t + \sqrt{2 h y}(1+y) \mathrm{d} W .
\end{equation}
Agreement between agent-based model and stochastic model for $y$ is demonstrated in Fig. \ref{fig:sdeAbmKirman}.

Note that the above derivation, and thus, the final equations,
does not change even if $\sigma_1$, $\sigma_2$ or $h$ are
functions of $x$ or $y$. Thus, one can further study the
possibilities of the model by checking different scenarios of
$\sigma_1$, $\sigma_2$ or $h$ being functions of $x$ or $y$.
Nevertheless, the most natural way is to introduce a custom
function $\tau(y)$ as inter-event time. In such case the switching
probabilities above can be interpreted as probability fluxes per
time unit. And thus, one can divide the aforementioned constants
by $\tau(y)$. We have chosen the case of
\begin{equation}
\mathrm{d} y = \left( \sigma_1 + y \frac{2 h - \sigma_2}{\tau(y)}
\right) (1+y) \mathrm{d} t + \sqrt{\frac{2 h y}{\tau(y)}} (1+y)
\mathrm{d} W , \label{eq:ytaucommon}
\end{equation}
as nonlinear SDE driving statistics of return in financial market.
We did not devide $\sigma_1$ by $\tau(y)$ on purpose as one could
argue that individual behavior of fundamentalist trader does not
depend on the observed returns.

\begin{figure*}[!t]
\centerline{ \subfloat{\includegraphics[width=2.5in]{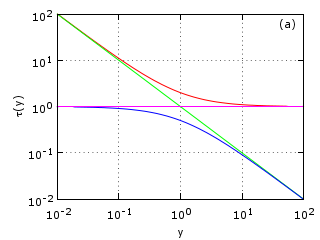}}
\hfil \subfloat{\includegraphics[width=2.5in]{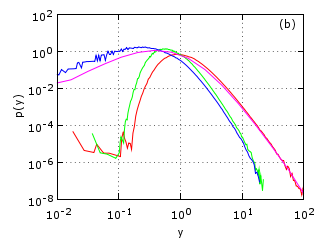}}
} \centerline{
\subfloat{\includegraphics[width=2.5in]{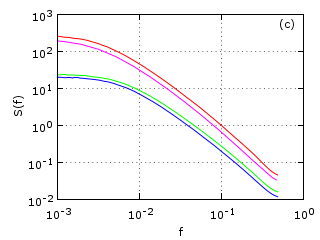}} }
\caption{Statistical properties, probability density functions
(sub-figure (b)) and power spectral density (sub-figure (c)),
obtained while solving (\ref{eq:ytaucommon}) with different
$\tau(y)$ scenarios (sub-figure (a)) being in use. Model
parameters were set as follows: $\sigma_{1}=\sigma_{2}=0.009$,
$h_0=0.003$.} \label{fig:tauScen}
\end{figure*}

In Fig. \ref{fig:tauScen}, we have shown statistical properties of
the stochastic model (\ref{eq:ytaucommon}) with different
$\tau(y)$ scenarios in use.

Note that while obtained stochastic model appears to be too crude
to reproduce statistical properties of financial markets in such
details  as our stochastic model \cite{GontisScyio2010} based on
empirical analyzes, it  contains long range power-law statistics
of return. Obtained equations are very similar to some general
stochastic models of the financial markets \cite{Ruseckas2010,
Reimman2010} and thus, in future development might be able to
serve as microscopic justification for them and maybe for our more
sophisticated modeling \cite{GontisScyio2010}.

\section{Two treatments of the Bass Diffusion model}
The Bass model introduces a differential equation for the diffusion rate of new products or technologies \cite{Bass1969}
\begin{IEEEeqnarray}{l l l}
\label{eq:bass}
\frac{\mathrm{d} N(t)}{\mathrm{d} t} & = & [ M - N(t) ] [p + \frac{q}{M} N(t) ] ,\\
N(0) & = & 0 .
\end{IEEEeqnarray}
where $N(t)$ denotes the number of product users at time $t$; $M$ is a market potential (number of potential users), $p$ is the coefficient of innovation, the likelihood of an individual to adopt the product due to influence by the commercials or similar external sources, $q$ is the coefficient of imitation, a measure of likelihood that an individual will adopt the product due to influence by other people who already adopted the product. This nonlinear differential equation serves as a macroscopic description of new product adoption by customers widely used in business planning \cite{Mahajan1993}.

Another approach to the same problem is related with agent based modeling of product adoption by individual users,
 or agents. The diffusion process is simulated by computers, where individual decisions of adoption occur with specific
  adoption probability affected by the other individuals in the
  neighborhood. It is easy to show that Bass diffusion process is
  a specific case of Kirman's herding model \cite{Kirman1993}. Indeed,
  lets define $x(t)$ as $x(t)=N(t)/M$ and in analogy with Kirman's model probability
  that new user will adopt the product as
\begin{equation}
\pi^{+}(x) = (1-x) \left(\frac{\sigma_1}{M} + \frac{h}{M} x
\right). \label{eq:pibass}
\end{equation}
In the case of Bass diffusion process is of one direction and
$\pi^{-}(x)=0$. Note that we assume an extensive herding in
equation (\ref{eq:pibass}) as only in this case the stochastic
term in corresponding Langevin equation vanishes with
$M\rightarrow \infty$. Then the functions defining the macroscopic
system description are as follows
\begin{IEEEeqnarray}{l l l}
A(x) &=& M \pi^{+}(x)= (1-x) \left(\sigma_1 + h x
\right) ,\\
D(x) &=& \pi^{+}(x)= \frac{(1-x)}{M} \left(\sigma_1 + h x \right).
\end{IEEEeqnarray}
In the limit $M\rightarrow \infty$ one gets Bass diffusion
equation (\ref{eq:bass}) with $p=\sigma_1$ and $q=h$ instead of
Langevin equation. This proofs that Bass diffusion is a special
case of Kirman's herding model. Though this simple relation looks
straightforward, we derive it and confirm by numerical simulations
in fairly original way.

\begin{figure*}[!t]
\centerline{ \subfloat{\includegraphics[width=2.5in]{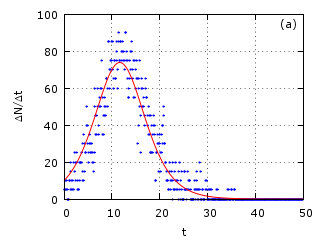}}
\hfil \subfloat{\includegraphics[width=2.5in]{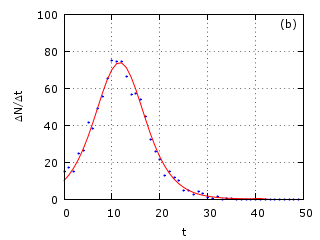}} }
\centerline{ \subfloat{\includegraphics[width=2.5in]{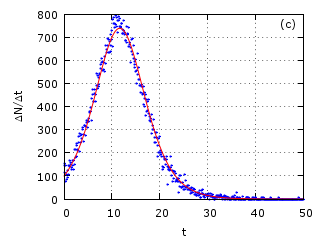}}
\hfil \subfloat{\includegraphics[width=2.5in]{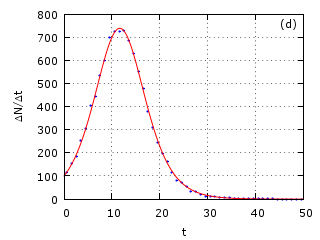}} }
\caption{Comparison of Bass diffusion, $\Delta N/\Delta t$ versus
$t$, in macroscopic description (red line) and agent based model
(blue points) shows convergence when time interval $\Delta t$ or
number of potential users $M$ are increasing. (a) $M=1000$,
$\Delta t=0.1$; (b) $M=1000$, $\Delta t=1$; (c) $M=10000$, $\Delta
t=0.1$; (d) $M=10000$, $\Delta t=1$. Other parameters are as
follows $p = 0.01$, $q = 0.275$.} \label{fig:bassAbmSde}
\end{figure*}

In Figure \ref{fig:bassAbmSde} we demonstrate the correspondence
between macroscopic and microscopic Bass diffusion description.
Agent based and continuous descriptions of product adoption
$\Delta N$ per time interval $\Delta t$ converge when number of
potential users $M$ or time interval $\Delta t$ increase.

One of the goals of developing these models on the web site
\cite{RizikosFizikaBusiness} was to provide theoretical background
of Bass diffusion model and practical steps how such computer
simulations can be created even with limited IT knowledge and then
used for practical purposes. Thus, we target small and medium
enterprises to encourage them to use modern computer simulation
tools for business planning and other purposes.

Computer models published at the \cite{RizikosFizikaBusiness}
provide a relatively easy starting point to get acquainted with
computer simulation and enables portal visitors to use these
computer models interactively, running them directly in a window
of web browser, changing parameters and observing results. This
significantly increases accessibility and dissemination of these
simulations.

\section{Conclusions and future work}
Reasoning of stochastic models of complex systems by the
microscopic interactions of agents is still a challenge for
researchers. Only very general models such as Kirman's herding
model in ant colony  or Bass diffusion model for new product
adoption have well established agent based versions and can be
described by stochastic or ordinary differential equations. There
are many different attempts  of microscopic modeling in more
sophisticated systems, such as financial markets or other social
systems,  intended to reproduce the same empirically defined
properties. The ambiguity of microscopic description in complex
systems is an objective obstacle for quantitative modeling. Simple
enough agent based models with established or expected
corresponding macroscopic description are indispensable in
modeling of more sophisticated systems. In this contribution we
discussed  various extensions and applications of Kirman's herding
model.

First of all, we modify Kirman's model introducing interevent time
$\tau(y)$ or trading activity $1/\tau(y)$ as functions of driving
return $y$. This produces the feedback from macroscopic variables
on the rate of microscopic processes and strong nonlinearity in
stochastic differential equations responsible for the long range
power-law statics of financial variables. We do expect further
development of this approach introducing the mood of chartists as
independent agent based process.

One more outcome of Kirman's herding behavior of agents is one
direction process - Bass diffusion. This simple example of
correspondence between very well established microscopic and
macroscopic modeling becomes valuable for further description of
diffusion in social systems. Models presented on the interactive
web site \cite{RizikosFizika} have to facilitate further extensive
use of computer modeling in economics, business and education.

\newpage
\section*{Acknowledgment}

Work presented in this paper is supported by EU SF Project
``Science for Business and Society'', project number:
VP2-1.4-\={U}M-03-K-01-019.

We also express deep gratitude to Lithuanian Business Support Agency.

%\IEEEtriggeratref{0}
\bibliographystyle{IEEEtran}
\bibliography{kirmanbass}

\end{document}